\documentclass[pra,showpacs,twocolumn]{revtex4}
\usepackage{amsmath}
\usepackage{bm}
\usepackage[dvips]{graphicx}
\usepackage{epsfig}

\begin{document}

\title{Stochastic theory of quantum vortex on a sphere}
\author{Hiroshi Kuratsuji}
\address{Research Organization of Science and Engineering, Ritsumeikan University-BKC,
              Kusatsu City, 525-8577, Japan }

\date{\today}

\begin{abstract}
A stochastic theory is presented for a quantum vortex that is expected to 
occur in superfluids coated on two dimensional sphere $ {\rm S}^2 $. The starting point is   
the canonical equation of motion ( the Kirchhoff equation) for a point vortex, 
which is derived using the time-dependent Landau-Ginzburg theory.  The vortex equation, 
which is equivalent to the spin equation, turns out to be the Langevin equation, 
from which the Fokker-Planck equation is obtained by using the functional integral technique.  
The Fokker-Planck equation is solved for several typical cases of the vortex motion 
by noting the specific form of pinning potential. 
An extension  to the non-spherical vortices is briefly discussed for the 
case of the vortex on plane and  pseudo-sphere.
\end{abstract}


\maketitle

\maketitle

\section{Introduction}
    
The study of vortex dynamics  has been one of central subjects in classical physics\cite{Lamb}.  
The modern interest of vortex phenomena is motivated by the quantum fluids,  
namely, the quantum vortex in various contexts,  which are considered to be defects inherent 
in the complex order parameters describing the quantum fluids.  Specifically, the dynamics of point vortex has 
been extensively investigated so far \cite{Onsager,Fetter,Chiao}. The dynamics of vortex string has been 
also studied \cite{Regge}.  Besides the conventional vortex in superfluid He4, the more complicated textural 
structure in He3 has been explored (\cite{Ho}).  Furthermore even the vortex in 
the cosmological scale has been explored(e.g.\cite{Kibble}) . 

In conventional treatment the vortex is an object defined on Euclidean space
of 2 and/or 3 dimension.  The vortex defined on curved space rather than flat Euclidean space 
is possible to occur.   Indeed, the classical counterpart of non-planer vortex has been given 
in several contexts starting with the purely classical hydrodynamical procedure,see e.g. 
\cite{Kimura,Kidambi}.  

On the other hand, it is known that in condensed matter physics the vortex motion is 
influenced by randomness caused by e.g. the temperature fluctuation and the presence 
of the interaction with the various sorts  of  impurities.  For such cases, we need to treat it within a 
framework of stochastic theory.   Indeed, the stochastic theory of vortex has been investigated 
in connection with the superconductivity \cite{Dorsey} .   

The purpose of this note is to present  a general formalism for stochastic theory of quantum  vortex that is 
constrained on the two dimensional manifold; specifically, 
two dimensional sphere $ {\rm S}^2 $. The essential point  is to follow an analogy between 
vortex and ferromagnetic spin: that is realized for a single magnetic ``domain" 
\cite{Brown}.     In order to  achieve this,  we first derive 
the equation of motion for the vortex center on these manifold; the canonical  (Kirchhoff) equation 
by using the time-dependent 
Landau-Ginzburg theory \cite{Kura}, that is described by the complex order parameter on $ {\rm S} ^2 $
(Section 2) .  Then the equation of motion for the vortex is converted to 
the Langevin equation by noting the analogy between the spherical vortex and 
the spin in the presence of the random force (Section 3). From this analogy,  
the Fokker-Planck equation is derived by adopting the functional 
integral technique. The functional integral used here is borrowed from 
the procedure that was developed for the statistical theory of wave propagation in random media\cite{Fibich,Langer}.  In section 4, we show the solutions of  the FP equation for a typical 
form of {\it pinning potential}.   In the section 5,  we give a brief sketch 
of an extension of the spherical vortex to its variants; the planer vortex and then pseudo-sphere vortex.


\section{Equation of motion for quantum vortex}

We start with the time-dependent Landau-Ginzburg Lagrangian, that is expressed in 
terms of the order parameter $ \psi $: 
\begin{equation}
L = \int [\frac{i\hbar}{2}(\psi^{*}\frac{\partial \psi}{\partial t}- c.c.)- {\cal H}(\psi. \psi^{*})]d\sigma .
\end{equation}
This form is the same as that was used for the superfluid which is defined 
on a flat plane\cite{Kura} . Here $ \psi ({\bf r}, t) $ is defined on 2-sphere $ S^2  $ 
with radius $ \vert {\bf r} \vert = a $ and the integral $ \int d\sigma $  
is taken over $ S^2 $, the coordinate of which is denoted by $ {\bf r} $ written in terms of 
the Cartesian form.  $ {\mathcal H} $ represents the Hamiltonian density which consists of 
the kinetic and potential energy;  
\begin{equation}
 {\cal H} = \frac{\hbar^2}{2m}\nabla\psi^{*}\nabla\psi + V(\psi^{*}, \psi) , 
\end{equation}
where $ \nabla $ means the derivative with respect to the spherical coordinate and $ m $ means the 
mass of constituent particles (e.g. He4 atoms).  

Let us consider the case that a single vortex is created, for which 
the order parameter is written by a form such that 
\begin{equation}
\psi ({\bf r}, t) = \psi ({\bf r}-{\bf R}(t)) . 
\end{equation}
Here $ {\bf R} $ denotes the time-dependent vortex center satisfying $ \vert {\bf R} \vert = a $. 
With this parametric form we have 
\begin{equation}
\frac{i\hbar}{2}(\psi^{*}\frac{\partial \psi}{\partial t}- c.c.)
= -\frac{i\hbar}{2}(\psi^{*}\nabla_{r} \psi - c.c.)\frac{d{\bf R}}{dt} , 
\end{equation}
where use is made of the relation:
$$  \frac{\partial \psi}{\partial t} = \frac{d{\bf R}}{dt}\nabla_R\psi 
$$
together with the relation  $ \nabla_R = - \nabla_r $ (that is, the 
the nabla with respect to the vortex center coordinate tuns out to be the 
derivative with respect to field argument $ {\bf r} $). 
Now we adopt the polar form of  the order parameter 
\begin{equation}
\psi = \sqrt{\rho}\exp[i\frac{m}{\hbar}\alpha({\bf r}-{\bf R}(t))] . 
\end{equation}
Here $ \rho= \rho({\bf r}-{\bf R}(t)) $ is the density, which vanishes at the vortex center $ {\bf r} = {\bf R} $ and 
tends to constant value: $ \rho = \rho_0 $ outside the coherent length.  The first term of $ L $ (denoted by $ L_c $ that is called the canonical term) is calculated as 
\begin{equation}
L_c = \int {\bf j}\cdot \frac{d{\bf R}}{dt}d\sigma 
\equiv  \int (j_x\cdot \dot X + j_y\cdot \dot Y + j_z\cdot \dot Z) d\sigma, 
\end{equation}
where {\it mass current } $ {\bf j} = m\rho\nabla \alpha = m\rho{\bf v}  $ is defined. 
On the other hand, the Hamiltonian term becomes 
\begin{equation}
H = \int [\frac{1}{2}m\rho{\bf v}^2 + 
 \frac{1}{4\rho}(\nabla\rho)^2 + V(\rho)]d\sigma
\label{Hamilton} 
\end{equation}
The first term means the fluid kinetic energy and the second and third terms 
are the internal energy, which is quoted as $ U(\rho) $. The explicit form will 
be given later. 

Now the equation of motion for the vortex center is derived from the Euler-Lagrange equation: 
$$
\frac{d}{dt}\big(\frac{\partial L}{\partial\dot{\bf R}}\big) - \frac{\partial L}{\partial{\bf R}} = 0 . 
$$
The contribution from the Hamiltonian is 
$$
\frac{\partial L}{\partial {\bf R}} = \frac{\partial}{\partial {\bf R}}\int {\mathcal H} d\sigma
= \frac{\partial H }{\partial {\bf R}} . 
$$
On the other hand, the canonical term is manipulated as follows: First by differentiating the under integral 
symbol, we have 
$$  \frac{d}{dt}\big(\frac{\partial L_c}{\partial\dot X}\big) 
= \int  \frac{d j_x}{dt}d\sigma
$$
together with $  \frac{d j_x}{dt} = \frac{\partial j_x}{\partial X}\dot X 
+ \frac{\partial j_x}{\partial Y}\dot Y 
+ \frac{\partial j_x}{\partial Z}\dot Z $
and 
$$  \frac{\partial L_c}{\partial X} = 
\int (\frac{\partial j_x}{\partial X}\dot X 
+ \frac{\partial j_y}{\partial X}\dot Y 
+ \frac{\partial j_z}{\partial X}\dot Z)d\sigma . 
$$
Hence we obtain
\begin{equation}
\frac{d}{dt}\big(\frac{\partial L_c}{\partial\dot {\bf R}}\big) - \frac{\partial L_c}{\partial {\bf R}}
= \int \{(\nabla \times {\bf j})\times   \dot {\bf R} \}d\sigma , 
\label{canonical}
\end{equation}
where use is made of $  \frac{\partial {\bf j}}{\partial X}  
= - \frac{\partial {\bf j}}{\partial x}  $.  
By evaluating the RHS of (\ref{canonical}) (see Appendix A),  
 we arrive at the Kirchhoff equation of motion for the vortex center 
\begin{equation}
\mu \Omega (\hat{\bf R}  \times \frac{d\hat{\bf R}}{dt}) = \frac{\partial H}{\partial \hat {\bf R}}
\label{vortex}
\end{equation}
with $ \Omega = m\rho_0 a^2 $ and replacing as $ H \rightarrow H/a $.  $ \mu $ means 
the ``vortex charge", which is introduced a multiplied factor in velocity field (Appendix A). 
Eq.(\ref{vortex}) coincides with the equation of motion for classical vortex on sphere 
\cite{Kidambi,Kimura}, which is derived in the framework of classical fluid mechanics.  
This equation has a simple 
dynamical meaning; the balance of two types of forces; the left hand side represents 
the Magnus force and the right hand side is the gradient force coming from the 
Hamiltonian (that plays a role of potential energy in analogy with classical particle mechanics). 
This equation of motion can be written in an alternative form: 
\begin{equation}
\mu\Omega\frac{d{\bf \hat R}}{dt} = - {\bf \hat R} \times  \frac{\partial H}{\partial {\bf \hat R}}  ~. 
\label{Kirch}
\end{equation}

The equation of motion is rewritten in a form of Hamiltonian equation of motion. 
Namely, by using the component for spherical vector basis 
$$
  \frac{d\hat{\bf R}}{dt} = (0, \dot\Theta, \sin\Theta \dot \Phi) , 
$$
the equation of motion is written in terms of the angular form: 
\begin{equation} 
\mu\Omega \dot\Theta  =  \frac{1}{\sin\Theta}\frac{\partial H}{\partial \Phi}, 
~\mu \Omega \dot\Phi   = -\frac{1}{\sin\Theta}\frac{\partial H}{\partial \Theta} . 
\end{equation}
This is of the same form as the equation of motion for spin (see, e.g. \cite{Suzuki,Mizobuchi}). 
The effective action corresponding to the equation of motion may be 
given as 
\begin{equation}
S =  \int [\mu\Omega(1-\cos\Theta)\dot\Phi - H]dt . 
\end{equation}

{\it Quantization of vortex charge}:   
We examine a special feature of the above form of action function; namely, 
the line integral. As a particular case, we consider the integral along a closed loop, say, $ C $. 
This can be written by the surface surrounding $ C $; where there is ambiguity 
choosing it, 
\begin{eqnarray}
\int_C \mu \Omega(1-\cos\Theta) d\Phi & = & \mu\Omega \int_S \sin\Theta d\Theta d\Phi \nonumber \\ 
                                                                       & = & - \mu\Omega \int_{\hat S} \sin\Theta d\Theta d\Phi 
\end{eqnarray}
which consists of the upper or lower surfaces, respectively, say  $ S $ and $ \hat S $ 
that are complement each other; $ S+ \hat S = S^2 $.  This ambiguity is related to the gauge choice
\cite{Sakurai}. Upon quantization, the ambiguity is solved; that is expressed as an exponentiation; 
$$ \exp[\frac{i}{\hbar}\mu\Omega \int_S \sin\Theta d\Theta d\Phi]
= \exp[-\frac{i}{\hbar}\mu\Omega \int_{\hat S} \sin\Theta d\Theta d\Phi] 
$$
then we obtain
$$ \mu  \Omega \int_{S^2} \sin\Theta d\Theta \wedge d\Phi = 2n\pi\hbar 
~~ ({\rm n} = {\rm integer} )
$$ 
which leads  to the Dirac monopole quantization:
\begin{equation}
4\pi \mu \Omega \equiv \mu M = \hbar \frac{n}{2}, 
\label{Dirac}
 \end{equation}
where $ M = 4\pi\Omega $, which means the mass of fluid on the sphere. 
Equation (\ref{Dirac}) is just the same as the relation between electric and 
magnetic charge for the case of quantized monopole.  

{\it Hamiltonian term; pinning potential}:  

The Hamiltonian term is evaluated as follows:  First we note that the fluid 
kinetic energy (\ref{Hamilton}) together with the second term which is given by 
the gradient of the density profile.  These two are independent of the position of vortex 
center for the case of single  vortex.  We need to settle a coordinate that is relative to the vortex center.  
Then the term satisfying this criterion is the last term $ V(\rho) $.   
In order to express this explicitly,  we consider the interaction energy with the pinning centers 
that is built in the superfluid, which is written in a form 
\begin{equation}
U= \int \psi^{*}({\bf r})\psi({\bf r})V({\bf r}) d{\bf r}
\end{equation}
and noting that the density profile is given as 
$ \vert \psi({\bf r}) \vert^2 = \rho_0 + \tilde\rho(\vert {\bf r} - {\bf R} \vert )  $ 
where $ \rho_0 $ is the uniform background term and 
$ \tilde\rho $ describes the vortex profile, which may be given by a Gaussian form 
with the central peak at the vortex center. The  interaction $ V({\bf r}) $ can be 
chosen such that it couples in a contact form; namely, that may be given 
by the delta function: 
\begin{equation}
V({\bf r} ) = V_0\delta({\bf r} - {\bf a}) 
\end{equation}
with the pinning center $ {\bf a} $, then we have $ U = V_0 \tilde\rho({\bf R} - {\bf a}) $ 
up to an additional constant term coming from the uniform background. 
If we take $ {\bf a} = (0,0,a)  $ as the north-pole: it turns out to be 
\begin{equation}
U = V_0\tilde\rho(a\sqrt{\vert (1-\cos\Theta\vert} ) 
\label{potential}
\end{equation}
where $ \Theta $ denotes the angle between the vortex center and the north-pole. 
In the case that there are several pinning centers located at  $ {\bf a}_k $, U is given by 
$$  U = \sum_{k=1}^{N} V_{0k}(\tilde\rho(\vert{\bf R} - {\bf a}_k\vert ) 
$$

\section{The Langevin and Fokker-Planck equations}

In this section we consider the stochastic equation for the vortex motion 
by following the analogy between spin equation of  motion.

\subsection{Spin analogy and the  functional integral}

As is remarked in the previous section, the equation of motion (\ref{Kirch}) is of same form as the spin 
equation of motion; the vector $ {\bf \hat R} $ plays a role for the spin vector, which we call the 
{\it pseudo-spin}. Here we introduce $ {\bf J} = J {\bf \hat R} $, and $ 
J \equiv \mu\Omega$ corresponds to the spin magnitude.  In terms of $ {\bf J} $, the above equation 
is written in a scaled form: 
\begin{equation}
\frac{d{\bf J}}{dt} = - {\bf J} \times  \frac{\partial H}{\partial {\bf J}}  ~. 
\label{spin}
\end{equation}
The right hand side of the equation represents the torque, namely, 
$ \frac{\partial H}{\partial {\bf J}} $ is nothing but the {\it magnetic field} acting on the 
pseudo-spin, which may be called ``pseudo-magnetic field". 
From physical point of view, the analogy between spin and vortex is not 
surprising, since the vortex may be regarded as a {\it region} where 
the angular momentum of the fluid is concentrated. 

Furthermore, following the well known fact in the spin theory,  
it can be extended to include the effect of dissipation: 
such that $ \nabla H $ is replaced by $ \nabla H + \eta \frac{d{\bf J}}{dt} $ in (\ref{spin}); 
with $ \eta $, the dissipative coefficient.  Having solved the equation for $ \frac{d{\bf J}}{dt} $, 
it follows (which is known as `` Landau-Lifschitz equation" \cite{Landau}): 
\begin{equation}
  \frac{d{\bf J}}{dt} +  {\bf A(J)} =  0  ~, 
 \label{dissip1}
\end{equation}
where we use the notational convention: 
\begin{equation}
{\bf A}({\bf J})  = \frac{1}{1+\eta^2J^2}\big[{\bf J }\times \frac{\partial H}{\partial {\bf J}} - 
 \eta J^2 \frac{\partial H}{\partial {\bf J}}\big],  
\end{equation}

We now treat the Brownian motion of point vortex that is caused 
by  random effects of several origins; the temperature fluctuations, and 
inevitably existing impurities. Here we follow an analogy that is expected to occur 
between a single vortex and a ferromagnetic spin which can be observed in a single 
ferromagnetic domain \cite{Brown}. For the case of ferromagnetic spin, the simplification 
is adopted; the random thermal fluctuations have a correlation time much shorter than the response 
time of the system. The response time of a single domain particle is of the same order as the 
reciprocal of the gyro-magnetic resonance frequency.  However, the case of the vortex on a sphere, 
it is not feasible  to estimate  the ratio between two characteristic times, so we adopt it as a 
working hypothesis.  

In what follows we consider that the fluctuation effect  gives  rise to the 
randomness of the pseudo-magnetic field; which may 
be written as $ {\bf b} $, so we replace $ \frac{\partial H}{\partial {\bf J}} \rightarrow 
\frac{\partial H}{\partial {\bf J}} + {\bf b} $.  Here ${\bf b}(t)$ is assumed to be the Gaussian white noise as a random uncorrelated magnetic field.  Hence we have 
\begin{equation}
\frac{d{\bf J}}{dt} +  {\bf A}({\bf J})   =  {\bf c}(t)  ~, 
 \label{dissip2}
\end{equation}
where ${\bf c}(t)$ the right-hand side, that is
\begin{equation}
{\bf c}(t)  = \frac{1}{1+ \eta^2J^2}({\bf J} \times {\bf b}-\eta J^2{\bf b}) 
\end{equation}
which is a combination of random force and torque.  As a result of of the random 
uncorrelated features of the function ${\bf b}  $,  we assume that  ${\bf c} $ can also be 
expected to be Gaussian white noise, which is expressed as 
\begin{eqnarray}
\langle c_i(t) \rangle & = & 0 ~, \nonumber \\
\langle c_i(t)c_j(t+u) \rangle & = & h \delta_{i,j} \delta( u) ~. 
\nonumber 
\end{eqnarray}
with $\delta(u)$ the delta function. It should be noted that the validity of the above form of white noise is not easy 
to justify and it is nothing else than the working hypothesis. 
The white noise is introduced to express  that the random magnetic field is correlated
on time-scales much smaller than the characteristic response time of the 
pseudo-spin system.
It is assumed that  $\langle {\bf c}^2 \rangle = 2h $, and 
its probability distribution may be given by the standard Gaussian 
functional form \cite{Fibich,Langer}: 
\begin{equation}
 P[{\bf c}(t)] = \exp \left[- \frac{1}{2h}\int_0^t  {\bf c}^2(t)dt \right] 
 \end{equation}
Using this distribution, the propagator $K$ between two definite spin states
at two different times, is given by the functional integral:
\begin{eqnarray}
K[{\bf J}(t)\vert {\bf J}(0)]  & = & \int \prod_{t} \delta[\frac{d{\bf J}}{dt} 
+ {\bf A}({\bf J}(t)) - {\bf c}(t)]  
\nonumber \\
& \times & 
\exp \left[- \int \frac{{\bf c}^2(t)}{2h}dt \right] \mathcal{D}[{\bf J}] \mathcal{D}[{\bf c}(t)]
\label{PI1}
\end{eqnarray}
with $\delta$ being the Dirac delta- functional.  We need to calculate the Gaussian integral with 
respect to $ {\bf c} (t) $, in which the functional Jacobian factor should be carried out. The details of 
this is briefly given in Appendix B and the result is 
\begin{equation}
K[{\bf J}(t)\vert {\bf J}(0)] = \int  \exp 
\left[- \frac{1}{2h}\int_0^t  \left(\frac{d{\bf J}}{dt} + {\bf A(J)} \right)^2dt \right] 
\mathcal{D}[{\bf J}]
\label{PI2}
\end{equation}
Expanding the squared term inside the exponential term, one can put this functional integral  
in a familiar form of the path integral for a particle in the vector potential $ \bf A $ and 
the scalar potential $ V = \frac{{\bf A}^2}{2} $ . 
The parameter $ h $ just corresponds to the Planck constant. 
We can formally write the above functional integral by the quantum mechanical 
path integral,  that is,  by using the imaginary time $ \tau = it $ 
\begin{equation}
K = \int \exp[\frac{i}{h}\int \{\frac{1}{2}\left(\frac{d{\bf J}}{d\tau}\right)^2 + i{\bf A}\cdot \frac{d{\bf J}}{d\tau} 
-  V\}d\tau ] {\mathcal D}{\bf J}(t)
\label{Feynman}
\end{equation}

\subsection{The Fokker-Planck equation}

The derivation of the Fokker-Planck is derived most directly by 
using the above path integral (\cite{Feynman}).  If we introduce 
the ``wave function"  $ \psi({\bf J}, \tau) $, we have the integral equation:
\begin{equation}
\Psi({\bf J}, \tau) = \int K[{\bf J}(\tau)\vert {\bf J}(0)] \Psi({\bf J}, 0) d{\bf J}(0)
\end{equation}
and following the standard procedure, we obtain the Schroedinger equation:
\begin{equation}
ih\frac{\partial \Psi}{\partial \tau} = \left[\frac{1}{2}({\bf P} +i{\bf A})^2 + V\right]\Psi 
\nonumber 
\end{equation}
with $ {\bf P} = -ih\frac{\partial}{\partial {\bf J}} \equiv -ih\nabla  $. Returning to the real time; namely, 
$ i\frac{\partial }{\partial \tau} = - \frac{\partial }{\partial t} $: then writing $ \Psi \rightarrow  P $: 
we arrive at the standard form of the FP equation\cite{Kubo}: 
\begin{equation}
\frac{\partial P}{\partial t} = h\nabla^2P - \nabla\cdot ({\bf A}P) 
\end{equation}
This can be rewritten as the continuity equation that is written as 
$ \frac{\partial P}{\partial t} + \nabla\cdot {\bf s} = 0 $, 
where $ {\bf s } $ denotes the probability current $ {\bf s } = -h\nabla P + {\bf A}P $ and 
the components in polar coordinate $ (\Theta, \Phi) $ are   
\begin{eqnarray}
s_{\theta} & =&  -h\frac{\partial P}{\partial \Theta} + \frac{P}{1+\eta^2J^2} \left(\eta J^2\frac{\partial H}
{\partial \Theta} 
-  \frac{1}{\sin\Theta}\frac{\partial H}{\partial \Phi}\right)   \nonumber \\
s_{\phi} & = &   -\frac{h}{\sin\Theta}\frac{\partial P}{\partial \Phi} + \frac{P}{1+\eta^2J^2} 
\left(\frac{\partial H}{\partial \Theta} 
+ \frac{\eta J^2}{\sin\Theta}\frac{\partial H}{\partial \Phi}\right) \nonumber 
\end{eqnarray}
and the FP equation is written as 
\begin{equation}
\frac{\partial P}{\partial t} = - \frac{1}{\sin\Theta}\big\{\frac{\partial}{\partial \Theta}(\sin\theta s_{\theta}) 
+ \frac{\partial s_{\phi}}{\partial \Phi}\big\} 
\end{equation}

We here consider several general consequences from the FP equation. 

(i): {\it The stationary distribution}:  We consider  $ \frac{\partial P}{\partial t} = 0 $, and we put 
ansatz of the Boltzmann distribution: $  P({\bf J}) 
= \exp[-\beta H] $ with $ \beta $ the inverse temperature $ \beta = \frac{1}{k_BT} $. 
By substituting this into the RHS of the FP equation, we get 
\begin{equation}
(\beta h - \frac{\eta J^2}{1 + \eta^2J^2}) \{\nabla^2 H - \beta(\nabla H)^2\} =0 
\nonumber 
\end{equation}
From this it follows the relation:
\begin{equation}
\beta h - \frac{\eta J^2}{1 + \eta^2J^2} = 0 
\label{FD}
\end{equation}
This relation is just the fluctuation dissipation relation, which establishes the 
relation between the dissipation coefficient $ \eta $ and the diffusion (fluctuation) 
coefficient $ h $.  

(ii): {\it Evolution equation for average value}\cite{Dyson}: 
The FP equation enables us to evaluate the average of functions on the 
sphere; $ F({\bf J}) $:  that is given by 
$ \langle F \rangle = \int F({\bf J}) P({\bf J}) d{\bf J} $: 
\begin{equation}
\frac{\partial \langle F \rangle }{\partial t} = h \langle \nabla^2F\rangle 
 - \langle \nabla F\cdot {\bf A}\rangle  
\label{average}
\end{equation}
where use is made of partial integration. Here  as an interesting example, we take up 
$ F({\bf J}) = J_i^2 (i =x,y,z) $ .  Further we choose $ {\bf A} = \gamma {\bf J} $, which 
represents the relaxation effect, so we have for $ \langle F \rangle $, 
\begin{equation}
\frac{d\langle J_i^2 \rangle}{dt} = 2h - \gamma \langle J_i^2  \rangle
\end{equation}
This is solved as 
\begin{equation}
\langle J_i^2  \rangle = \frac{2h}{\gamma}(1 -\exp[-\frac{\gamma}{2h}t])
\end{equation}
Namely, this shows a typical relaxation behavior of  $ \langle J_i^2 $ leading to 
the asymptotic value $  \langle J_i^2(\infty) \rangle  =  \frac{2h}{\gamma} $.

\section{Simple examples: the effect of pinning potential}

The FP equation is of peculiar form that is somewhat different from the usual one that consists of the 
potential term, so we need to invoke specific techniques to deal with this. We now consider  
an example that can be treated by analytic as well as approximate ways. 

As the Hamiltonian we take the type given by the form such that 
it is given as a function of $ \Theta $: That is given by the pinning potential  (\ref{potential}) 
which comes from the pinning center located at the north pole. As a concrete form 
we consider the case that the profile $ \tilde \rho $ is given by  the Gaussian \cite{note}: 
$ \tilde\rho = \exp[ - \frac{({\bf r}- {\bf R})^2}{\alpha}] $ with the vortex size $ \alpha $, 
hence we have 
\begin{equation}
H(\Theta) = V_0\exp[-a^2(1- \cos\Theta)/\alpha] \equiv  V_0\tilde H. 
\end{equation}
As another form, it is possible to adopt the potential arising from two pining centers; one is located 
at  the north and the other is at the south pole: and is written as 
\begin{equation}
H(\Theta) = V_0\big\{\exp[-\frac{a^2}{\alpha}(1- \cos\Theta)]  + \exp[-\frac{a^2}{\alpha}(1 + \cos\Theta)]\big\}
\nonumber 
\end{equation}

In the following we consider the two cases without or with the effect of dissipation.  

(i)  {\it The case that there is no dissipation: $ \eta = 0 $}: 
We examine this case in the semiclassical way based on the classical equation of motion; 
the ``Lagrangian"  in the (\ref{PI2}) is 
is 
\begin{equation}
{\mathcal L} = \frac{1}{2} J^2\big[ \dot\Theta^2 + \sin^2\Theta(\dot\Phi 
+ \frac{V_0}{J\sin\Theta}\frac{d\tilde H}{d\Theta}  )^2\big] 
\end{equation}
We here examine a special solution such that  $ \dot\Theta = 0 $, which leads to 
$ \Theta = \Theta_0 $.    Then, the functional integral is 
\begin{equation}
K = \int \exp[-\frac{1}{2h}\int_0^t \sin^2\Theta_0(\dot\Phi + \kappa)^2dt]
{\mathcal D}(\Phi)
\end{equation}
with 
$$ 
\kappa \equiv \big[\frac{V_0}{J\sin\Theta}\frac{d\tilde H}{d\Theta}\big]_{\Theta= \Theta_0}
$$
This is equivalent to the path integral for a particle on a circle; 
the corresponding FP equation becomes 
\begin{equation}
h\frac{\partial P}{\partial t} = \frac{h^2}{\sin^2\Theta_0}\frac{\partial^2P}{\partial \Phi^2}
  - h\kappa \frac{\partial P}{\partial \Phi}
\end{equation}
which is a modified form of the diffusion equation. 
We assume the solution to be a form 
\begin{equation}
P(\Phi,t) = \sum_{n= -\infty}^{\infty} \exp[-\lambda_n t] f_n(\Phi) 
\label{solution}
\end{equation}
Here $ f_n(\Phi) $ satisfies the eigenvalue equation
\begin{equation}
\frac{h}{\sin^2\Theta_0}\frac{d^2f_n}{d\Phi^2}
  - \kappa \frac{d f_n}{d\Phi} = -\lambda_n f_n 
\nonumber 
\end{equation}
We put  $ f_n(\Phi) = \exp[in\Phi] ~(n={\rm integers}) $, and 
the eigenvalue is obtained as 
$$ 
\lambda_n = \frac{h}{\sin^2\Theta_0}n^2 +  in\kappa 
$$
By substituting this into the (\ref{solution}), we have 
\begin{equation}
P(\Phi)  = \sum_{n= -\infty}^{\infty} \exp[- (\frac{h}{\sin^2\Theta_0}n^2 +  in\kappa)t] 
\exp[in\Phi]
\end{equation}
which turns out to be the theta function\cite{Toda}. The expectation values for 
$ \langle {\bf J}\rangle  
=  \int_0^{2\pi} {\bf J}  P(\Phi, t) d\Phi $ is calculated as 
\begin{eqnarray}
\langle J_x \rangle &= & J\sin\Theta_0\exp[-\frac{ht}{\sin\Theta_0}]\cos\kappa t \nonumber \\
 \langle J_y \rangle &= & = J\sin\Theta_0\exp[-\frac{ht}{\sin\Theta_0}]\sin\kappa t \nonumber \\
 \langle J_z \rangle  & = &  J\cos\Theta_0 
 \end{eqnarray}
The damping behavior of $ (J_x, J_y ) $ shows is similar to the diffusive behavior 
of spin in magnetic systems \cite{Yoshida}.

\bigskip 

(ii)  {\it The case that there is dissipation: $ \eta \ne 0 $}.  

For this case, we look for the 
distribution function as a function of $ \Theta $ only $ P \equiv P(\Theta) $ , so the stationary 
FP equation is given by 
\begin{equation}
\frac{h}{\sin\Theta}\frac{d}{d\Theta}(\sin\Theta\frac{dP}{d\Theta}) 
- \frac{V_0\eta J^2}{1 + \eta^2J^2}\frac{1}{\sin\Theta} \frac{d}{d\Theta}(\sin\Theta\frac{d\tilde H}{d\Theta}P) 
= -E P \nonumber 
\end{equation}
which is rewritten as 
\begin{equation}
\frac{1}{\sin\Theta}\frac{d}{d\Theta}(\sin\Theta\frac{dP}{d\Theta}) 
- \frac{\epsilon}{\sin\Theta} \frac{d}{d\Theta}(\sin\Theta\frac{d\tilde H}{d\Theta}P) 
= -\lambda P 
\end{equation}
Here we put $ \epsilon = \frac{V_0\eta J^2}{h(1 + \eta^2J^2)} $  and 
$ \lambda = \frac{E}{h} $. This can be regarded as an expansion parameter. Noting that 
$ \epsilon \equiv \frac{V_0\eta J^2}{h(1 + \eta^2J^2)} $ turns out to be 
$ \epsilon = V_0\beta $ as a consequence of the fluctuation-dissipation relation 
(\ref{FD}), we can treat the expansion scheme in terms of the expansion 
parameter $ \epsilon $  

(a):  We first consider the perturbation procedure: this may be relevant for the case that 
$ V_0\beta <<1 $, which means the temperature is higher enough the interaction 
strength satisfies $ V_0 << k_BT $.  Hence expanding in terms of power series with respect to $ \epsilon $: 
\begin{eqnarray} 
P(\Theta) & = & P^{(0)}(\Theta) + P^{(1)}(\Theta) + \cdots \nonumber \\
\lambda & = & \lambda_0 + \lambda_1 + \cdots  \nonumber 
\end{eqnarray}
Substituting this into the above eigenvalue equation, we can determine 
each order by iteration procedure: the zero-th order 
\begin{equation}
\frac{h}{\sin\Theta}\frac{d}{d\Theta}(\sin\Theta\frac{dP^{(0)}}{d\Theta}) = 
 -\lambda_0P^{(0)}
 \end{equation}
 and the first order:  
 \begin{eqnarray} 
\frac{h}{\sin\Theta}\frac{d}{d\Theta}(\sin\Theta\frac{dP^{(1)}}{d\Theta})  
  -     \frac{\epsilon}{\sin\Theta} \frac{d}{d\Theta}(\sin\Theta\frac{d\tilde H}{d\Theta})P^{(0)}  & & \nonumber \\
  = 
   -\lambda_1P^{(0)} - \lambda_0P^{(1)}.
\end{eqnarray}
The zeroth order gives the Legendre polynomial: 
$ P^0 = P_l(\Theta) $ and the corresponding eigenvalue $ \lambda_0 =  l(l+1) $ 
with $ l = 1 \cdots  $ . Let us choose a particular $ l $, namely, $ P^0 = P_l(\Theta) $ ; then 
the first order term is calculated to 
\begin{equation}
\lambda_1 = \epsilon 
\int P_l(\Theta) \frac{1}{\sin\Theta}\frac{d}{d\Theta}(\sin\Theta\frac{d\tilde H}{d\Theta})P_l(\Theta)d\Omega/ 
\int P^2_l(\Theta)d\Omega 
\end{equation}
$ P^{(1)} $ can be expanded in terms of $ P_k(\Theta) $:
$  P^{(1)} = \sum_k C_k P_k(\Theta) $, then we obtain  the coefficients $ C_k $ 
\begin{equation}
C_k =  \frac{V_{kl}}
{\lambda_0(l) - \lambda_0(k)} 
\end{equation}
with the matrix element: 
$$ V_{kl} \equiv \epsilon \int P_k(\Theta)  \frac{1}{\sin\Theta}\frac{d}{d\Theta}(\sin\Theta\frac{d\tilde H}{d\Theta})P_l(\Theta)d\Omega
$$
for $ l \neq k $.  Using this, he second order shift of the eigenvalue is calculated 
\begin{equation}
\lambda_2 = \epsilon^2 \sum_{k\neq l} \frac{V_{kl}^2}{\lambda_0(l) - \lambda_0(k)} 
\end{equation}

(b):  We next examine the case that $ \epsilon \simeq 1 $, namely, $ V_0 \simeq k_BT $. In this case one has to use 
non-perturbative procedure \cite{Brown}. In order to carry out this, we rewrite  the FP equation 
in terms of the variable $ \xi = \cos\Theta $:  
\begin{equation}
\frac{d}{d\xi}\big[(1-\xi^2)\big\{\frac{dP}{d\xi}
+ \epsilon\frac{d\tilde H}{d\xi}P\big\}\big] = -\lambda P
\nonumber 
\end{equation}
and if noting the relation 
\begin{equation}
\exp[-\epsilon \tilde H]\frac{d}{d\xi}\exp[\epsilon\tilde H] 
= \frac{d}{d\xi} + \epsilon\frac{d\tilde H}{d\xi}  
\nonumber 
\end{equation}
we have 
\begin{equation}
\frac{d}{d\xi}\big\{(1-\xi^2) \exp[-\epsilon \tilde H]\frac{d}{d\xi}\big(\exp[\epsilon \tilde H]P\big)\big\}
= -\lambda P
\nonumber 
\end{equation}
This can be further reduced by putting $ Q = \exp[\epsilon \tilde H]P $, 
\begin{equation}
\frac{d}{d\xi}\big\{(1-\xi^2) \exp[-\epsilon \tilde H]\frac{dQ}{d\xi}\big\}
= -\lambda \exp[-\epsilon \tilde H]Q
\label{Q}
\end{equation}
This equation can be translated to the variational problem; namely, we
introduce the ``action"; 
\begin{equation}
I [Q] = \int^{1}_{-1} (1-\xi^2)\exp[-\epsilon \tilde H]\big(\frac{dQ}{d\xi}\big)^2 d\xi
\label{action}
\end{equation}
and  the constraint coming from the normalization; 
\begin{equation}
\int^{1}_{-1}\exp[-\epsilon \tilde H]Q^2 d\xi \equiv N[Q] =1
\label{constraint}
\end{equation}
Taking the ``Euler-Lagrange equation"  $ \delta (I[Q] - \lambda N[Q]) = 0 $, we 
recover the above eigenvalue equation with $ \lambda $ is nothing but 
the Lagrangian multiplier. We see that there is a  trivial solution $ Q_0 = $ constant  
with the zero eigenvalue $ \lambda_0  =0  $.  Starting from this as the lowest solution, we 
have an algorithm to get the subsequent solutions; $ Q_n (n=0, \cdots ) $ by noting the orthogonality 
condition; 
\begin{equation}
\int_{-1}^1 \exp[-\epsilon \tilde H]QQ_n d\xi = 0 
\label{ortho}
\end{equation}

As a concrete case, we exemplify how to construct  the second  lowest state 
by taking a trial function given by the quadratic function $ Q_1(\xi) = A + B\xi + C\xi^2 $. 
Substituting this into (\ref{ortho}), we have the linear relation between $ A, B,C $ and 
$ C $ can be expressed by $ A, B $.  Next  (\ref{constraint}) gives the quadratic 
equation for  $ A, B $. Finally  (\ref{action}) gives another quadratic form 
for $ A,B $.  Then, the problem is reduced to the minimization of the (\ref{action}) 
written in terms of the quadratic form  under the constraint  (\ref{constraint}) written by 
the quadratic equation. Having solved  these, we arrive at the solution for $ Q_1 $. 
This procedure may proceed to step by step and we can have a series of 
the stationary solution of the FP equations.

\section{Extension to non-spherical vortex}

Here we discuss that the non-spherical vortices can be readily obtained as variants 
of the vortex on the sphere that has been given above;  we are concerned with the vortex 
on 2-dimensional plane  and the vortex on pseudo-sphere. 
\subsection{The case of planer vortex}

This can be regarded as a limiting case of the spherical vortex.  
For the vortex center coordinate 
 $ {\bf R} = (X,Y) $,  the equation of motion for $ {\bf R} $ is  obtained as a special case 
 of the infinite radius; $ a \rightarrow \infty $; 
\begin{equation}
m\rho_0 \mu({\bf k}  \times \frac{d{\bf R}}{dt}) = \frac{\partial H }{\partial {\bf R}}
\end{equation}
where $ {\bf k} $ is the unit vector perpendicular to $ (x, y) $ plane. 
The Langevin equation with dissipation together with the white noise $ {\bf c} $ is given by 
\begin{equation}
\frac{d\bf R}{dt} + {\bf B}   =  {\bf c} 
\end{equation}
Following the same procedure as the case of spherical vortex, we obtain 
the FP equation: 
\begin{equation}
h\frac{\partial P}{\partial t} = \nabla^2P - \nabla\cdot({\bf B}P) 
\end{equation}
where $ \nabla = (\frac{\partial}{\partial X}, \frac{\partial}{\partial Y}) $ is just 
the nabla in  two dimensional plane. In the above, $ {\bf B} $ is 
\begin{equation}
{\bf B}  \equiv  \frac{1}{(m\rho_0\mu)^2 + \eta^2}(m\rho_0\mu {\bf k} \times \frac{\partial H}{\partial {\bf R}} 
+ \eta\frac{\partial H}{\partial {\bf R}} ) 
\nonumber 
\end{equation}
This is essentially same as that has been used in the case of superconductivity vortex \cite{Dorsey}. 
More detailed and complicated treatment was given for the actual superconductors in connection with 
high $ T_c $ superconductivity \cite{Feigelman,Enomoto}.  Our treatment here provide a general way 
to clarify the stochastic behavior of planer vortex. The details of the argument will be given elsewhere. 

\subsection{The case of vortex on pseudo-sphere}
 
Next  we  consider a point vortex on the pseudo-sphere (which is denoted by $ PS^2 $) which 
has the  surface with constant negative curvature. The superfluid 
coated on such pseudo-sphere may be realized in actual conditions 
as a local part of the superfluid that is put on the surface of complicated shape and 
this part can be regarded as a pseudo-sphere. 

The point of the surface is described by the equation $ x^2 + y^2 - z^2 = -a^2  $, 
which is parametrically 
written as  
$$  x= \sinh\theta\cos\phi, x= \sinh\theta\sin\phi, z = \cosh\theta
$$
Let  
 $ {\bf \hat R}  \equiv {\bf R}/a = (\sinh\Theta\cos\Phi, \sinh\Theta\sin\Phi, \cosh\Theta) $
 be the  vortex center, which satisfies\cite{Matsumoto}   
\begin{equation}
 \mu\Omega\frac{d{\bf \hat R}}{dt} = -{\bf \hat R} * \frac{\partial H}{\partial{\bf \hat R}}
\label{pseudo}
\end{equation}
Here the scalar and vector product , denoted as $ * $,  are  defined by 
\begin{eqnarray}
{\bf a}\cdot{\bf b} & = & a_1b_1 + a_2b_2 - a_3b_3 \nonumber \\
  {\bf a}*  {\bf b} & = & (a_2b_3 -a_3b_2,  a_3b_1 -a_2b_3,
-(a_1b_2 -a_2b_1))  
\nonumber 
\end{eqnarray}
and the triple product formula
$$
{\bf a}*({\bf b}*{\bf c}) = ({\bf a}\cdot {\bf b}){\bf c} 
- ({\bf a} \cdot {\bf c}){\bf b} 
$$
Note that the 3-rd component of the scalar product has a minus sign compared with the spherical case. 
Furthermore noted is that for the case of vortex on $ PS^2 $, there is no 
counterpart of the topological quantization, because $ PS^2 $ is trivial in a 
topological sense, namely, this is iso-morphic to 2-dimensional 
flat Euclidean space.  We introduce the pseudo spin; $ {\bf K} = K\hat {\bf R} $ with $ K \equiv \mu\Omega $. 
The norm of $ {\bf K} $ is given by $ {\bf K}^2 = -K^2 $. $ K $ is not quantized, which is a consequence 
of the non-existence of topological invariant for $ PS^2 $. 
According to the same manner as the sphere vortex, the Langevin equation with 
the dissipation is obtained as 
\begin{eqnarray}
 \frac{d{\bf {\bf K}}}{dt}  & = & \tilde A({\bf K})  + \tilde{\bf c} \nonumber \\
 \tilde A({\bf K}) & \equiv & \frac{1}{1- \eta^2K^2}
  (-{\bf  K} * \frac{\partial H}{\partial{\bf K}} - \eta K^2\frac{\partial H}{\partial {\bf K}}) 
\end{eqnarray}
Here note that  $ c $, which is the random force, has the norm such that $ {\bf \tilde c}^2 
= c_1^2 + c_2^2 - c_3^2 $, which is a consequence of the non-Euclidean metric 
on the pseudo-sphere.   By taking account of this fact and following the same procedure 
as the spherical vortex case, we arrive at the FP equation; 
\begin{equation}
\frac{\partial P}{\partial t} = - \frac{1}{\sinh\Theta}\big\{\frac{\partial}{\partial \Theta}(\sinh\theta s_{\theta}) 
+ \frac{\partial s_{\phi}}{\partial \Phi}\big\} 
\end{equation}
where the current is 
\begin{eqnarray}
s_{\theta} & =&  -h\frac{\partial P}{\partial \Theta} - \frac{P}{1-\eta^2K^2} \left(\eta K^2\frac{\partial H}
{\partial \Theta} 
-  \frac{1}{\sinh\Theta}\frac{\partial H}{\partial \Phi}\right)   \nonumber \\
s_{\phi} & = &   -\frac{h}{\sinh\Theta}\frac{\partial P}{\partial \Phi} + \frac{P}{1-\eta^2K^2} 
\left(\frac{\partial H}{\partial \Theta} 
- \frac{\eta K^2}{\sinh\Theta}\frac{\partial H}{\partial \Phi}\right) \nonumber 
\end{eqnarray}

We can develop the approximate procedure in the same way as the spherical vortex. 
This may be realized by finding the pinning potential, which can be obtained 
by replacing $ \cos\Theta \rightarrow \cosh\Theta $ (\ref{potential}).  
For developing the perturbation, the zeroth order (unperturbed) distribution function 
satisfies $ \Theta $, 
\begin{equation}
\frac{1}{\sinh\Theta}\frac{\partial}{\partial\Theta}(\sinh\Theta \frac{\partial P}{\partial \Theta})
= \lambda P 
\nonumber 
\end{equation}
and the eigenvalue is given by $ \lambda = \rho(\rho +1) $ with $ \rho $ is the 
real number and the eigenfunction is written in terms of the integral representation
\cite{Helgason}: 
\begin{equation}
P_{\rho}(\cosh\Theta) = \frac{1}{2\pi}\int_0^{2\pi} (\cosh\Theta + \sinh\Theta \cos\xi)^{\rho}d\xi 
\nonumber 
\end{equation}
Further, for the non-perturbation case, we can construct the similar equation 
with (\ref{Q}) as well as the (\ref{action}), which is simply obtained by replacing 
$ \xi =\cos\Theta $ by $ \xi = \cosh\Theta $.  Then we can apply the variation principle 
for $ I[Q]  $ leading to  the approximation solutions for the FP equation.

\section{Summary}

We have presented a general formalism of the stochastic theory of random behavior of the 
quantum vortex that is coated on the two-dimensional sphere.  The randomness 
are caused from several origins.  We have mind of the conventional assumption; 
temperature fluctuations.  For the case of mesoscopic size we have in mind, 
the fluctuation of temperature is  more appreciable than 
the macroscopic system from the relative sense.  As the other origin, we 
can take into account of the one that is intrinsically built in the spherical geometry, 
for which we expect to have the irregular deviation from the ideal sphere in 
actual situation\cite{Turner,Bowick}.  

To formulate the random behavior, we follow the analogy with the spherical vortex and spin. 
This analogy seems natural; since the vortex carries angular momentum of the fluid in a form 
of ``lump" which is the small region that angular momentum is concentrated. 
From the technical point of view of stochastic theory, the present 
approach is closely connected with the magnetic system\cite{Brown}.
Having noted the analogy with spin, we can  write the Langevin equation for the 
Brownian motion of vortex, which can be converted to the Fokker-Planck equation 
using the functional integral method\cite{Fibich,Langer}.  Though this present approach shares 
its idea with the FP equation for magnetic system\cite{Brown}, the present 
functional integral provides with the more direct means to access to the 
FP equation, that is quite different  from the stochastic integral method (Ito integral) 
used in \cite{Brown}.  As a particular case, the approximate scheme has been developed 
for the case that the vortex Hamiltonian is given by the pinning potential. 
This case may be significant for practical analyses, since the pinning plays 
a crucial role for the vortex motion.  

Finally, we gave an extension to the case of the variants of the spherical vortex; 
the planer vortex and the vortex on the pseudo-sphere. 
The vortex on the pseudo-sphere may be regarded as a step towards the quantum 
vortex that is defined on the more complicated two-dimensional manifold.


\centerline{\bf Acknowledgement}

The author would like to thank professor R.Botet at Laboratoire de Physique Solides,  Orsay 
for useful discussions on the 
Fokker-Planck equation.

\begin{appendix}

\section{Calculation of $ \nabla\times {\bf j} $} 

We rewrite the surface integral of the right hand side of (\ref{canonical}) 
by using the Stokes theorem: 
\begin{equation}
\int_S (\nabla \times {\bf j})  d\sigma = \int_C m\rho {\bf v} d{\bf s}  
\label{vorticity}
\end{equation}
with $ C $ being boundary of S . By noting the fact that 
if $ C $ is chosen such that it is sufficiently remote from the vortex center, 
$ \rho \rightarrow \rho_0 $,  we have  
$$
\int_C m\rho {\bf v} d{\bf s} = m\rho_0 \int_C {\bf v} \cdot d{\bf s}
\equiv m\rho_0 \int_S {\bf \omega} d\sigma 
$$
with the vorticity $ {\bf \omega} = \nabla \times {\bf v} $. 

\bigskip

Next we give an explicit form for the velocity field, which is given by  
\begin{equation}
{\bf v} = \mu (\hat r \times \nabla f) , 
\end{equation}
where the {\it vortex charge} $ \mu $ is introduced and  $ f $ is
\begin{equation}
f = \log \vert {\bf r} - {\bf R}(t) \vert = \log a + \log \vert \hat{\bf r} - \hat{\bf R}(t) \vert . 
\end{equation}
Here $ \hat{\bf r} = {\bf r}/a, \hat{\bf R}= {\bf R}/a $, that is , the vector on 
unit sphere: 
$$
 \hat R = (\sin\Theta \cos\Phi, \sin\Theta\sin\Phi, \cos\Theta) 
$$
 At this point we adopt the polar coordinate system: 
$ \hat{\bf r}= (\sin\theta\cos\phi, \sin\theta\sin\phi, \cos\theta) $ and 
the velocity field becomes  
\begin{equation} 
{\bf v} = \mu(0, -\frac{1}{\sin\theta}\frac{\partial f}{\partial \phi}, \frac{\partial f}{\partial \theta})
\end{equation}
Hence the vorticity is given by $ {\bf \omega} = \nabla \times {\bf v} = (\nabla^2 f)\hat r  $
with the Laplacian on 2-sphere
$$
\nabla^2 f = \frac{1}{\sin\theta}\frac{\partial}{\partial\theta}(\sin\theta \frac{\partial f}{\partial \theta} ) + 
\frac{1}{\sin^2\theta}\frac{\partial^2 f}{\partial \phi^2}= \delta (\hat{\bf r} - \hat{\bf R}(t))  . 
$$
where we note $  \nabla^2 \log \vert \hat{\bf r} - \hat{\bf R}(t)\vert 
=\delta (\hat{\bf r} - \hat{\bf R}(t)) $. From the above expression 
the vorticity for the spherical vortex directs to the normal of sphere. 
By substituting this into the Lagrangian equation of motion, we arrive at the 
Kirchhoff equation. 

{\it Vortex equation  on a pseudo-sphere}

Here we give a sketch for the vortex on pseudo-sphere $ PS^2 $ 

The procedure of calculation is completely same as the case of $ S^2 $; we need only 
the following changes: 
$$ \nabla f = (0, \frac{\partial f}{\partial\theta}, \frac{1}{\sinh\theta}\frac{\partial f}{\partial \phi}) 
\equiv  \frac{\partial f}{\partial\theta}\hat \theta + \frac{1}{\sinh\theta}\frac{\partial f}{\partial \phi}\hat \phi 
$$
where $ \hat\theta, \hat\phi $ means the unit vector in the direction of $ (\theta, \phi) $. 
Noting that the velocity field is given by replacing $ sin\theta $ by $ \sinh\theta $ and 
 hence the vortex becomes $ \omega = \nabla \times {\bf v} = (\nabla^2 f)\hat r $
with the Laplacian on pseudo-sphere
$$
\nabla^2 f = \frac{1}{\sinh\theta}\frac{\partial}{\partial}((\sinh\theta \frac{\partial f}{\partial \theta} ) + 
\frac{1}{\sinh^2\theta}\frac{\partial^2 f}{\partial \phi^2}
$$
Hence, we obtain the equation of motion in the same form as the spherical case 
by noting the difference of vector product.

\section{Reduction of functional integral}

Here we show how to reduce the Gaussian functional integral. 
We introduce $ {\bf G} = \frac{d{\bf J}}{dt} + {\bf A(J)} $.  First we use the functional identity:
\begin{equation}
\det \big[\frac{\delta{\bf G}(t)}{\delta {\bf J}(t)}\big]_{G=c} 
\int \prod_t \delta[{\bf G}({\bf J}) - {\bf c}] D[{\bf J}(t)]   = 1
\label{delta}
\end{equation}
then we have the functional integral including the 
the Jacobian factor\cite{Faddeev,Zinn}: 
\begin{eqnarray}
K & = & \int  \det \big[\frac{\delta {\bf G}(t)}{\delta{\bf J}(t)} \big]_{{\bf G=c}}
\big]\int  \prod_t \delta
[{\bf G}(t) -  {\bf c}(t)]D[{\bf J}(t)] \nonumber \\
& \times & 
 \exp[- \frac{D}{2}\int {\bf c}^2(t)dt] D[{\bf c}(t)]
 \end{eqnarray}
Here  we can remove  the constraint that attaches to the $ \det $, 
namely, $ {\bf G= c} $. Because, it can be obtained by the well known 
the relation for the delta function $ c(x)\delta(x-a) = c(a)\delta(x-a) $
and the determinant factor is generally regarded as a functional of $ {\bf J}(t) $, so
it does not depends on  $ {\bf c}(t) $ any more.  Further, using an integral identity 
$ \delta[g(x)] = \int \exp[i\lambda g(x)]d\lambda $, the above is written as a 
functional integration  over three variables $ {\bf  J}, {\bf c}, \bf\lambda $: 
Thus we have 
 \begin{eqnarray}
 K & = & \int \exp[i\int  \lambda(t)\cdot \{{\bf G}(t) - {\bf c}(t)\}dt]
  \nonumber \\
& \times  & 
\exp[- \frac{D}{2}\int {\bf c}^2(t)dt]D[\lambda(t)]
\nonumber \\
& \times & 
\det \big[\frac{\delta{\bf G}(t)}{\delta {\bf J}(t)}\big]
D[{\bf c}(t)]
D[{\bf J}(t)]
\label{func}
\end{eqnarray}
By accomplishing the two Gaussian functional integrals for 
$ {\bf c} $ and $ \lambda(t)  $ successively,  we obtain 
(\ref{PI2}). 

\end{appendix}


\begin{thebibliography}{10}
\bibitem{Lamb}
H.Lamb, {\it Hydrodynamics}
Cambridge University Press, Cambridge, 1935.
\bibitem{Onsager}
L.Onsager, Nouvo Cimento {\bf 69}, 279(1949) .
\bibitem{Fetter}
A.L.Fetter, Phys.Rev.{\bf 162},143(1967). 
\bibitem{Chiao}
R.Y.Chiao, A.Hansen, and A.A. Moulthrop, \hfill\break
Phys.Lett.Lett.
{\bf 54},1339(1985).
\bibitem{Regge}
M.Rasetti , and T.Regge, Physica {\bf 80A},217(1975). 
 \bibitem{Ho}
D.Mermin  and T.L.Ho, Phys. Rev.Lett. {\bf 36}(1976)594. 
\bibitem{Kibble}
T W B Kibble, J. Phys. {\bf A9},1387(1976) .   
\bibitem{Kimura} 
Y.Kimura and H.Okamoto, J.Phys.Soc.Jpn. \hfill\break
{\bf 56},4203(1987). 
 \bibitem{Kidambi}
R.Kidambi and P.K.Newton, Physica {\bf D140},95(2000).
\bibitem{Dorsey}
A.T.Dorsey, Phys.Rev. {\bf B46},)8376(1992. 
\bibitem{Brown}
W.F.Brown, Phys.Rev.{\bf 130},1677(1963). 
\bibitem{Kura}
H.Kuratsuji, Phys.Rev.Lett. {\bf 68},1746(1992).
 \bibitem{Fibich}
M.Fibich and E.Helfand, Phys.Rev.{\bf 183},265(1969).
\bibitem{Langer}
J.Zittarz and J.S. Langer, Phys.Rev.{\bf 148},741(1966).
%
\bibitem{Suzuki}
H.Kuratsuji and T.Suzuki, J.Math.Phys.{\bf 21},472(1980).
\bibitem{Mizobuchi}
H.Kuratsuji and Y.Mizobuchi, Phys.Lett.{\bf 82A}, \hfill\break
279(1981).
\bibitem{Sakurai}
J.J.Sakurai, {\it Modern Quantum Mechanics, second edition}, Addision Wesley, New York, 1995.
\bibitem{Landau}  L.Landau and E.Lifschitz, Phys.Z.Sovietunion {\bf 8},153 (1935).
\bibitem{Feynman}
R.Feynman and A.R.Hibbs, {\it Quantum Mechanics and Path Integral}, 
Macgraw-Hill, New York, 1965.   
\bibitem{Kubo}
R.Kubo, M.Toda  and N.Hashitsume, {\it Statistical Physics II}, 2-nd edition, Springer Verlag, 
New York, 1985. 
\bibitem{note}
Alternatively we can choose 
$ H = V_0\vert 1 -\cos\Theta \vert  ~({\rm for} ~\vert 1 -\cos\Theta \vert <\alpha ) $ 
and $ H = 0 $ for otherwise. 
\bibitem{Dyson}
F.Dyson, J.Math.Phys.{\bf 3}, 1191 (1962). 
\bibitem{Toda}
M.Toda, {\it Introduction to elliptic functions}(in Japanese), Nihon Hyoronsha, Tokyo, 2004.
\bibitem{Yoshida}
K.Yoshida, {\it Magnetism} (in Japanese),Iwanami Shoten, Tokyo, 2008. 
\bibitem{Feigelman}
M.V.Feigelman and V.M.Vinokur, Phys.Rev.{\bf B41}, 896(1990). 
\bibitem{Enomoto}
Y.Enomoto, Phys.Lett.{\bf A161}, 185 (1991). 
\bibitem{Matsumoto}
This equation of motion is similar to the one for the equation of motion of 
pseudo-spin on the non-compact phase space; $ SU(1,1)/U(1) $: 
See, M.Matsumoto and H.Kuratsui, Phys. Lett. {\bf A195}, 18(1994). 
\bibitem{Helgason}
S.Helgason, {\it Differential geometry and symmetric space}, Academic press, New York, 1962. 
\bibitem{Turner}
A.M.Turner, V.Vitelli and D.R.Nelson,  Rev. Mod. Phys. 82, 1301(2010) 
\bibitem{Bowick}
M.J.Bowick and L.Gioni, Adv.Phys.{\bf 58},449 (2009) and references 
therein.
\bibitem{Faddeev}
L.Faddeev and V.Popov, Phys.Lett.{\bf 25B}, 29 (1967).
\bibitem{Zinn}
J.Zinn-Justin, {\it Quantum field theory and critical phenomena}, 
Oxford University press, Oxford, 1990.  
\end{thebibliography}
\end{document}